\documentclass[runningheads]{llncs}
\usepackage[T1]{fontenc}
\usepackage{graphicx}
\usepackage{adjustbox}
\usepackage{caption}
\usepackage{subcaption}
\usepackage{amsmath}
\usepackage{amssymb}
\usepackage{times}
\usepackage{epsfig}
\usepackage{multirow}
\usepackage{hyperref}
\usepackage{color}
\begin{document}

%
%\title{Contribution Title\thanks{Supported by organization x.}}
\title{ECG-ATK-GAN: Robustness against Adversarial Attacks on ECGs using Conditional Generative Adversarial Networks}
\titlerunning{ECG-ATK-GAN}
% If the paper title is too long for the running head, you can set
% an abbreviated paper title here
%
%\author{Anonymous}
\author{
Khondker Fariha Hossain\inst{1} \and
Sharif Amit Kamran\inst{1} \and
Alireza Tavakkoli\inst{1} \and \\
Xingjun Ma\inst{2}}

%\index{Hossain, Khondker Fariha}
%\index{Kamran, Sharif Amit}
%\index{Tavakkoli, Alireza}
%\index{Ma, Xingjun}}

%
\authorrunning{Hossain et al.}
% First names are abbreviated in the running head.
% If there are more than two authors, 'et al.' is used.
%
\institute{Dept. of Computer Science \& Engineering, University of Nevada, Reno, NV, USA\and School of Computer Science, Fudan University, China}
%\institute{Anonymous organization\\
%\email{***@*****.***}\\}
%

\maketitle              % typeset the header of the contribution
\begin{abstract}
Automating arrhythmia detection from ECG requires a robust and trusted system that retains high accuracy under electrical disturbances. Many machine learning approaches have reached human-level performance in classifying arrhythmia from ECGs. However, these architectures are vulnerable to adversarial attacks, which can misclassify ECG signals by decreasing the model's accuracy. Adversarial attacks are small crafted perturbations injected in the original data which manifest the out-of-distribution shifts in signal to misclassify the correct class. Thus, security concerns arise for false hospitalization and insurance fraud abusing these perturbations. To mitigate this problem, we introduce the first novel Conditional Generative Adversarial Network (GAN), robust against adversarial attacked ECG signals and retaining high accuracy. Our architecture integrates a new class-weighted objective function for adversarial perturbation identification and new blocks for discerning and combining out-of-distribution shifts in signals in the learning process for accurately classifying various arrhythmia types. Furthermore,  we benchmark our architecture on six different white and black-box attacks and compare them with other recently proposed arrhythmia classification models on two publicly available ECG arrhythmia datasets. The experiment confirms that our model is more robust against such adversarial attacks for classifying arrhythmia with high accuracy.

\end{abstract}
\section{Introduction}
ECG is a crucial clinical measurement that encodes and identifies severe electrical disturbances like cardiac arrhythmia and myocardial infractions. Many artificial intelligence and machine learning approaches have been proposed to detect different types of ECGs accurately \cite{jambukia2015classification,mahajan2017cardiac,faziludeen2013ecg}. Recently, deep convolutional neural networks (CNNs) \cite{shaker2020generalization,acharya2017deep,kachuee2018ecg,ahmad2021ecg,9413938} has become the norm for achieving near-human-level performance for classifying cardiac arrhythmia and other cardiac abnormalities. Popular systems such as Medtronic LINQ II ICM \cite{medtronic}, iRhythm Zio \cite{irhythm}, and Apple Watch Series 4 \cite{han2020deep} use embedded DNN models to analyze cardiac irregularities by monitoring the signals. Accurately detecting arrhythmia in real-time enables an immediate referral of the patient to appropriate medical facilities. In addition to providing the patient with timely medical help, this will benefit the insurance companies by potentially reducing long-time consequences of delayed healthcare. Despite all these benefits, state-of-the art systems used to predict arrhythmia are vulnerable to adversarial attacks. These vulnerabilities are crucial as they can result in false hospitalization, misdiagnosis, patient data-privacy leaks, insurance fraud, and negative repercussion for healthcare companies \cite{finlayson2019adversarial,han2020deep}. 

Although these vulnerabilities are highly studied \cite{chen2020ecgadv,lam2020hard}, a comprehensive solution is yet to be devised. Adversarial attacks misclassify ECG signals by introducing small perturbations that inject the out-of-distribution signal into the classification path. The perturbations could be introduced to the data by accessing the model parameters (White-box attack) or inferring the bad prediction outputs for a given set of input (Black-box attack) \cite{chakraborty2021survey}. Current deep learning \cite{shaker2020generalization,acharya2017deep,kachuee2018ecg} and GAN-based \cite{hossain2021ecg,golany2019pgans,golany2021ecg,golany2020simgans} classifiers are not specifically designed to utilize the objective function to identify and mitigate adversarial attacked ECGs. Although recent works \cite{han2020deep,chen2020ecgadv} illustrated the vulnerability of deep learning architectures to adversarial attacks, our work proposes a first-of-its-kind defense strategy against six different adversarial attacks for ECGs using a novel conditional generative adversarial networks. Additionally, we incorporate a class-weighted categorical loss function for identifying out-of-distribution perturbations and emphasizing the class-specific features. Both qualitative and quantitative benchmarks on two publicly available ECG datasets illustrate our proposed method's robustness.

\begin{figure}[!tp]
    \centering
    \includegraphics[width=0.9\textwidth]{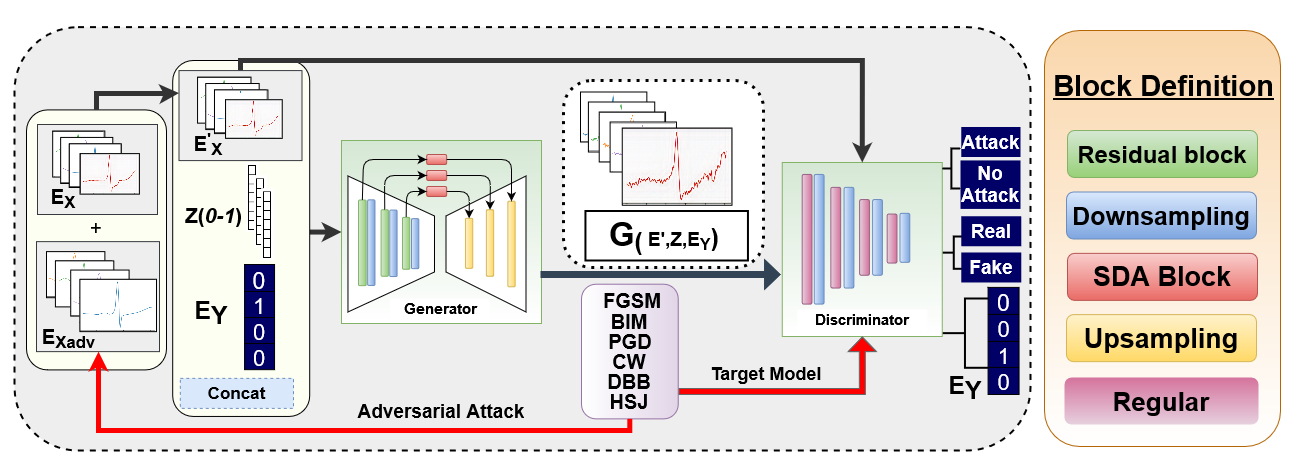}
    \caption{Proposed ECG-ATK-GAN consisitng of a Generator and a Discriminator. The discriminator is utilized for generating the six attacked signals  $E_{x_{adv}}$, namely FGSM, BIM, PGD, CW, HSJ, and DBB. These are then added with the non-attacked signals $E_x$ to create the training data-set $E^{'}_{x}$. Contrarily, the Generator takes both attacked and non-attacked ECG signals, $E^{'}_{x}$, a noise vector $z$, and the class labels $E_y$ as input. }
    \label{fig1}
\end{figure}

\section{Methodology}
\subsection{Generator and Discriminator}
\label{subsec:generators}
We propose a novel GAN based on a class-conditioned generator and a robust discriminator for categorical classification of both real and adversarial attacked ECG signals as illustrated in Fig.~\ref{fig1}. The generator concatenates both non-attacked or attacked ECG signals  $E_x'$, label $E_y$, and a noise vector $z$ as input and generates $G(E_x',Ey,z)$. We use a Gaussian filter with $\sigma=3$  to generate the smoothed noise vector, $z$. The label vector $E_y$ in our model is utilized so that generated signal is not random. Rather it imitates class-specific ECG representing an arrhythmia. The noise vector, $z$ ensures that the generated signal has small perturbations so that it does not fully imitate the original ECG signal and helps in overall training in extrapolation of generated signals. The generators incorporate Residual, Downsampling, Upsampling, and Skip-Dilated Attention (SDA) block as visualized in Fig.~\ref{fig1}.  \iffalse The residual and downsampling block's feature sizes are $[R1,R2,R3,D1,D2,D3]$= $[32,64,128]$,  the upsampling block's feature sizes are $[U1, U2, U3]$ =$[128,64,32]$ and the SDC block's feature sizes are $[S1, S2, S3]$ =$[32,64,128]$.\fi The generator uses Sigmoid activation as output, so the synthesized signal is constrained within 0-1 as a continuous value.

\begin{figure}[!tp]
    \centering
    \includegraphics[width=1\textwidth]{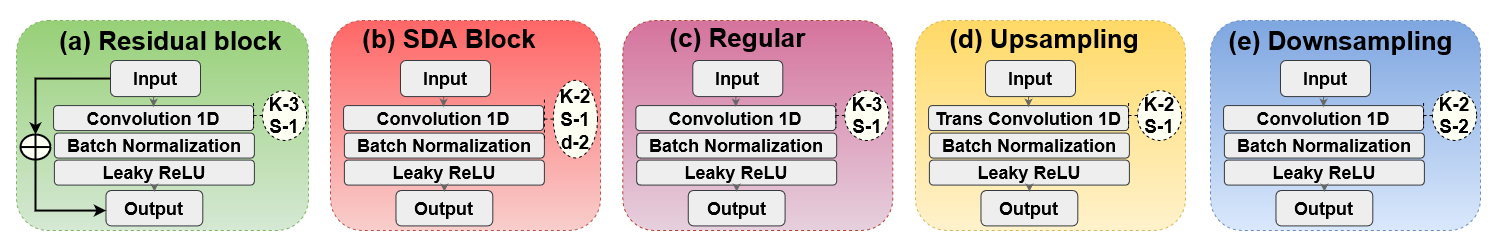}
    \caption{Proposed (a) Residual, (b) Skip-Dilated Attention, (c) Regular, (d) Upsampling and, (e) Downsampling Blocks. Here, K=Kernel size, S=Stride, and D=Dilation rate.}
    \label{fig2}
\end{figure}

The discriminator takes attacked/non-attacked real ECG, $x$ and GAN synthesized ECG, $G(E_x', E_y,z)$ signals sequentially while training. The discriminator consists of three regular blocks and three downsampling blocks (Fig.~\ref{fig1}). The discriminator utilizes three losses: 1) Class-weighted Categorical cross-entropy for identifying adversarial attacked/non-attacked ECGs, 2) Categorical cross-entropy for  normal and arrhythmia beat classification and 3) Mean-squared Error for GAN adversarial training. \iffalse We use downsampling blocks $3x$ times after each Des block. For convolution we use kernel size, $k=3$, and stride, $s=1$, except for downsampling blocks where we use stride, $s=2$.  The number of features are for Des and Downsampling blocks are  $[F1,F2,F3,D1,D2,D3]$= $[16,32,128,16,32,128]$. \fi So we use three output activations: Sigmoid (GAN training),  and two Softmax for adversarial attack and arrhythmia/normal beat classification.

\begin{itemize}
    \item \textbf{Residual Block:} For extracting small perturbations in the attacked ECG signals, we use convolution with a small kernel, $k=3$ and stride, $s=1$ in the residual block in the Generator, illustrated in Fig.~\ref{fig2}(a). This residual block is capable of extracting fine features that extrapolate the original signal to contain small perturbations and make it out-of-distribution. \iffalse As seen in Fig.~\ref{fig2}(a) the proposed residual block allows for the original unperturbed signal via its skip connection while producing negative samples through its main convolutional path that amplifies the perturbation with its Leaky-ReLU module. \fi Specifically, the residual skip connection retains important signal-specific information that is added with more robust features extracted after the batch-normalization and leaky-ReLU activation.

    \item \textbf{Skip-Dilated Attention Block:} We use skip-dilated attention (SDA) block with kernel size, $k=2$, dilation rate, $d=2$ and stride, $s=1$, as illustrated in Fig.~\ref{fig2}(b).  By utilizing dilated convolution, our receptive fields become larger, covering larger areas of the attacked signals \cite{yu2015multi}.\iffalse This extracts the direction of the small perturbations and mitigates the attack by reapplying the signal onto the original manifold. This is done specifically by combining the output of the SDA block with the output of the upsampling block. As a result more local and global signals are amplified in order to suppress the out-of-distribution perturbation \cite{zhang2019self,chen2018attention}. \fi
    \item \textbf{Regular Block:} We use the regular block for discriminators, containing convolution ($k=3, s=1$), batch-norm, and leaky-ReLU layers, as visualized in Fig.~\ref{fig2}(c). Our main objective here is to encode the signals to meaningful classification outputs for two tasks, which is to 1) classify the type of arrhythmia and, 2) distinguish between non-attacked/attacked signals. Therefore, we avoid using any complex block for feature learning and extraction.  
    \item \textbf{Downsampling and Upsampling Blocks:} The generator consist of both downsampling and upsampling blocks, whereas the discriminator consist of only downsampling blocks to get the desired feature maps and output.  The upsampling block consists of a transposed-convolution layer, batch-norm, and Leaky-ReLU activation layer successively and is given in  Fig.~\ref{fig2}(d). In contrast, The downsampling block comprises of a convolution layer, a batch-norm layer and a Leaky-ReLU activation function consecutively and is illustrated in Fig.~\ref{fig2}(e). 
\end{itemize}

\subsection{Objective Function and Individual Losses}
To distinguish non-attacked and attacked signals with out-of-distribution perturbations and emphasize the class-specific features even under significant perturbations, we propose a class-weighted categorical cross-entropy loss. The loss function is given in Eq.~\ref{eq1}, where $m =2$, for attacked/non-attacked signal and $\kappa$ is the class weight for the ground-truth, $E_y$ and predicted class-label, $E_{y'}$.
\begin{equation}
    \mathcal{L}_{atk}(D) =  -\sum^{m}_{i=0} \kappa^i E^i_y\log(E^i_{y'})
\label{eq1}
\end{equation}

For classification of normal and different arrhythmia signals, we use categorical cross-entropy loss. Here, $k=$  distinct normal/arrhythmia beats, depending on the dataset.

\begin{equation}
    \mathcal{L}_{ary}(D) =  -\sum^{k}_{i=0} E^i_y\log(E^i_{y'})
\label{eq2}
\end{equation}

For ensuring that the synthesized signal contains representative features of both adversarial examples and adversarial attacks, our generator incorporates the mean-squared error (MSE) as shown in Eq.~\ref{eq3}. This helps the generator output signals with small perturbations that guarantee the signal to misclassify. As the generator, $G$ is class-conditioned, it takes distinct ground truth class-label $E_y$, along with the attacked/non-attacked ECGs, $E^{'}_{x}$ and Gaussian noise vector $z$ as input.
\begin{equation}
    \mathcal{L}_{mse}(G) = \frac{1}{N} \sum_{i=1}^{N} (G(E^{'}_{x},E_y,z) - E_x )^2
    \label{eq3}
\end{equation}

We use Least-squared GAN \cite{mao2017least} for calculating the adversarial loss and training our GAN. The cost function for our adversarial loss is given in Eq.~\ref{eq4}. The discriminator takes real ECG signal, $E_{x}$ and generated ECG signal, $G(E^{'}_{x},E_{y},z)$ in two iterations. The adversarial loss quadratically penalizes the error while stabilizing the min-max game between the generator and discriminator. 
\begin{equation}
    \mathcal{L}_{adv}(D) =  \big[\ (D(E_x',E_y) -1)^2 \big]\ + \\ \big[\ (D(G(E_x',E_y,z),E_y)+1))^2 \big]\
\label{eq4}
\end{equation}
\iffalse
In Eq.~\ref{eq1}, we train the discriminators on the real ECG attacked and non-attacked signals, $E_x'$. Following that, we train with the generated attacked and non-attacked ECG signals, $G(E_x',E_y,z)$.  We begin by batch-wise training the discriminators $D$ on the training data. Following that, we train the $G$ while keeping the weights of the discriminators frozen. Likewise, we train $G$ on a batch of training samples while keeping the weights of all the discriminators frozen. \fi

By incorporating Eq.~\ref{eq1}, \ref{eq2}, \ref{eq3} and \ref{eq4}, we can formulate our final loss function as given in Eq.~\ref{eq5}. Here,  $\lambda_{mse}$, $\lambda_{atk}$, and $\lambda_{ary}$ denote different weights, that are multiplied with their corresponding losses. We want our generator to synthesize realistic ECGs to fool the Discriminator, while classifying the types of arrhythmia with high accuracy. So, the final goal is to maximize the adversarial loss and minimize other losses. \iffalse The weights decide which loss to prioritize while training. \fi
\begin{equation}
\min \limits_{G,D_{ary},D_{atk}} \big( \max \limits_{D_{adv}}  (\mathcal{L}_{adv}(D)) + \lambda_{mse}\big[\ \mathcal{L}_{mse}(G)\big]\ +  \\
\lambda_{atk}\big[\mathcal{L}_{atk}(D)\big]\ + \lambda_{ary}\big[\ \mathcal{L}_{ary}(D)\big]\ \big)    
\label{eq5}
\end{equation}

\begin{figure}[!tp]
    \centering
    \includegraphics[width=1\linewidth]{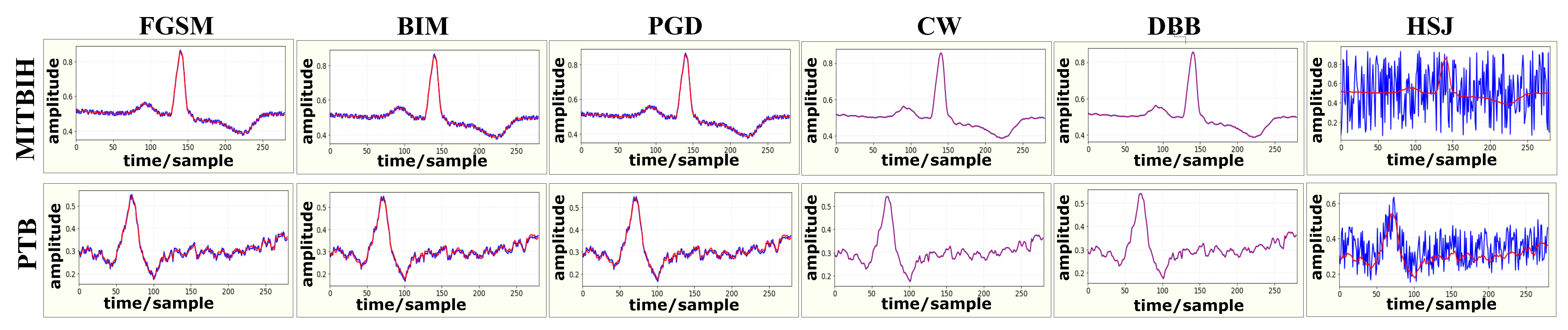}
    \caption{The non-attacked and attacked signals (white and black-box attacks) overlapped on each other signified by {\color{red} Red} and {\color{blue} Blue} lines. }
    \label{fig3}
\end{figure}

\subsection{Adversarial Attacks}
We incorporated six established adversarial attacks (shown in Fig~\ref{fig3}) that target our discriminator model as it is responsible for classifying different types of arrhythmia and normal beats in ECG signals. The reason for choosing these state-of-the-art attacks is to make our model more robust for intrusive perturbations in real-world applications. Four of these attacks are white-box, meaning detailed knowledge of the network architecture, the parameters, and the gradient w.r.t to the input is utilized to corrupt the data \cite{chakraborty2018adversarial}. The other two are black-box attacks, meaning no knowledge of the underlying architecture or parameter is needed; instead, some output is observed for some probed inputs \cite{chakraborty2018adversarial}. Moreover, the attack corrupts the data by estimating the gradient direction using the information at the decision boundary of the output \cite{chen2020hopskipjumpattack,brendel2017decision}. We experimented with perturbation values, $\epsilon$ ranging from $0.001$ to $0.1$, and selected the value which looked visually realistic and harder to discern.
So, the visually realistic perturbations for FGSM, BIM, PGD and DBB is, $\epsilon=0.01$ and for CW, $\epsilon=0.1$.
\begin{itemize}
    \item \textbf{Fast Gradient Sign Method (FGSM):}  This white-box attack creates attacked ECGs, $E_{X_{adv}}$ by perturbing the original signal, $E_{X}$. For this, it calculates the gradients of the loss, $\mathcal{L}_{ary}$ (Eq.~\ref{eq2}) based on the input signal to create new adversarial signals that maximize the loss \cite{goodfellow2014explaining}.
    \item \textbf{Basic Iterative Method (BIM):} This is an improved white-box attack, where the FGSM attack is iteratively updated in a smaller step size and clips the signals values of intermediate results to ensure the $\epsilon$-neighborhood of the original signal, $E_{X}$  \cite{kurakin2016adversarial}.
    \item \textbf{Projected Gradient Descent (PGD):} This white-box attack is considered the most decisive first-order attack. Though similar to BIM, it varies in initializing the example to a random point in the $\epsilon$-ball of interest (decided by the $L_\infty$ norm) and does random restarts. In contrast, BIM initializes in the original point \cite{madry2017towards}. 
    \item \textbf{Carlini-Wagner (CW):} This is an optimization-based white-box attack \cite{carlini2017towards}. It resolves the unboundedness issue by using line search to optimize the attack objective. We utilized the version with the $L_{\infty}$ norm, i.e., for maximum perturbation applied to each point in the signal. 
    \item \textbf{Decision-based Boundary Attack (DBB):} This is a decision-based black-box attack that starts from querying a large adversarial perturbation and then seeks to reduce the perturbation while staying adversarial \cite{brendel2017decision}. It only requires the final class prediction of the model.
    \item \textbf{Hop Skip Jump Attack (HSJ):} A powerful black-box attack that only requires the final class prediction of the model \cite{chen2020hopskipjumpattack}. And it is an advanced version of the boundary attack,  requiring significantly fewer model queries than Boundary Attack.
    
\end{itemize}

\section{Experiments}
\iffalse
In the following section, different experimentation and evaluation are provided for our proposed architecture. First, we elaborate on the data preparation in 3.1 and pre-processing scheme in Sec. 3.2. We then define our hyper-parameter settings in Sec. 3.3. Different architectures are compared based on some quantitative(Sec. 3.4) and qualitative (Sec. 3.5) evaluation metrics for real examples and adversarial examples. 
\fi
\subsection{Data Set Preparation}
We used the PhysioNet MIT-BIH Arrhythmia dataset for our experiment \cite{moody2001impact}. We divided the dataset into four categories, \textbf N [Normal beat, Left and right bundle branch block beats, Atrial and Nodal escape beat], \textbf S [Atrial premature beat, Aberrated atrial, Supraventricular and Nodal premature beat], \textbf V [Premature ventricular contraction, Ventricular escape beat], and \textbf F [Fusion of the ventricular and regular beat]. We first find the R-peak for every signal, use a sampling rate of 280 centering on R-peak, and then normalize the amplitude between $[0,1]$. In the benchmarking, we combine and split the samples into 80\% and 20\% sets of train and test data. So we end up having train samples of N: 69958, S: 4766, V: 1965, F:617, and test samples of N: 17571, S: 1126, V: 473, and F: 157. \iffalse As N has too many samples, we only use 10,000 out of 69958. \fi To overcome the lack of minority class samples, we use Synthetic Minority Over-sampling Technique (SMOTE)  \cite{chawla2002smote} to increase the number of samples for S, V, and F to 10,000 each. \iffalse, ending up with 40,000 training signals in total.\fi We do not use SMOTE on test data. Next, we use the train and test ECG signals to create the six types of adversarial attacked ECGs (using Adversarial Robustness toolbox \cite{art2018}). So we end up having \iffalse 40,000 training and 19,327 test\fi same number attacked ECGs as non-attacked ones for each adversarial attacks. Next, we combine the original and adversarial ECGs to create our whole  training dataset, $E_x + E_{xadv} = E^{'}_{x}$ (Fig.~\ref{fig1}). We use 5-fold cross validation and select the model with the best validation score. \iffalse of 80,000.\fi

We also benchmark on PTB Diagnostic ECG Database \cite{bousseljot1995nutzung}, which consists of Normal and Myocardial Infraction beats. \iffalse for our experiment. However, as there was a lack of samples for other classes, we did not use them in our experiment. \fi For each category, we use 10,000 samples, meaning we end up having 20,000 ECGs in total. We split them into 80\%  training and 20\% test data. In similar manner to MITBH, we apply  six adversarial attacks on these ECG signals. For training we end up having 32,000 (16,000 non-attacked and 16,0000 attacked) signals for each attack types. We use the same 5-fold cross-validation method.

\begin{table}[!tp]
\centering
\caption{\textbf{MIT-BIH Dataset} : Comparison of architectures trained and evaluated on \textbf{non-attacked/attacked} ECGs for normal and three arrhythmia beat classification.}
\begin{adjustbox}{width=1\linewidth,height=4.2cm}
\begin{tabular}{c|c|c|c|c|c|c|c|c|c|c}
\hline
\multicolumn{1}{l|}{\multirow{2}{*}{}} & \multirow{2}{*}{Model} & \multirow{2}{*}{Accuracy} & \multicolumn{2}{c|}{N}                              & \multicolumn{2}{c|}{S}                              & \multicolumn{2}{c|}{V}                              & \multicolumn{2}{c}{F}                              \\ \cline{4-11} 
\multicolumn{1}{l|}{}                  &          &            & \multicolumn{1}{c|}{Sensitivity} & \multicolumn{1}{c|}{Specificity}  & \multicolumn{1}{c|}{Sensitivity} & \multicolumn{1}{c|}{Specificity}   & \multicolumn{1}{c|}{Sensitivity} & \multicolumn{1}{c|}{Specificity}   & \multicolumn{1}{c|}{Sensitivity} & \multicolumn{1}{c}{Specificity} \\ \hline\hline
\multirow{4}{*}{\begin{tabular}[c]{@{}c@{}}No\\ Attack\end{tabular}}     &    \textbf{Proposed Method}      &    \textbf{99.2}      &    \textbf{98.8}    &     95.2  & 83.7 &  \textbf{99.8}       &     97.9       &   \textbf{99.7}      &   92.4      &         99.3  \\ \cline{2-11}
&     Shaker et al. \cite{shaker2020generalization}    &    98.6    &   97.4     &    \textbf{98.1}      &  \textbf{93.0} &    98.7          &  \textbf{99.2}           &     99.0                &       87.2        &      \textbf{99.6}       \\\cline{2-11}

&    Kachuee et al. \cite{kachuee2018ecg}    &      98.1    &   96.8      &     94.5                 & 88.7  &  97.6   &  92.5             &      99.6              &    90.4            &      99.3   \\ \cline{2-11}

&   Acharya et al. \cite{acharya2017deep}   &  96.4    &  92.8    &   96.2   &  86.2 & 97.0   &   95.9            &    98.8      &   \textbf{94.2}    &      97.1      \\ \hline\hline

\multirow{4}{*}{FGSM}     &    \textbf{Proposed Method}        &      \textbf{98.7}    &  \textbf{97.9}      &  \textbf{95.0}          &  \textbf{82.6} & \textbf{99.2}   &      \textbf{99.2}      &     \textbf{98.7}                      &   73.2       &   \textbf{99.8}      \\ \cline{2-11}

&       Shaker et al. \cite{shaker2020generalization}    &    92.6      &  84.7        &        93.3            & 81.8  & 89.9        &     96.5          &      95.6                      &          57.9      &      98.7       \\\cline{2-11}

&         Kachuee et al. \cite{kachuee2018ecg}    &       86.5   &  73.1     &     82.8                & 68.9 & 83.4   & 73.7         &      97.3       &   \textbf{82.8}                   &   93.2         \\ \cline{2-11}
&        Acharya et al. \cite{acharya2017deep}     &  77.2      &   53.3     &     87.9         &  65.7  &  74.1       &       66.2       &       92.3                  &  65.6    & 87.9          \\ \hline\hline
\multirow{4}{*}{BIM}     &   \textbf{Proposed Method}       &    \textbf{98.1}    &   \textbf{97.1}     &     \textbf{91.2}         &  \textbf{76.1} & \textbf{98.6}  &  95.0 & \textbf{99.4}  &  84.1        &     98.9              \\ \cline{2-11} 
&        Shaker et al. \cite{shaker2020generalization}    &  96.2       &  93.1    &   90.1          &  69.1 &   98.2         &   \textbf{97.6}    &    94.9                   &   55.4      &  \textbf{99.8}         \\ \cline{2-11} 
&         Kachuee et al. \cite{kachuee2018ecg}    &     85.6     &   70.6      &   90.4                  &  67.4 & 90.5    &    83.6           &         92.3             &    82.1        &  88.5           \\ \cline{2-11}
&        Acharya et al. \cite{acharya2017deep}     &   76.9     &  54.4     &   87.5        &  49.0  & 88.2      &   42.1            &        91.9                 &   \textbf{87.8}     &   73.8        \\ \hline\hline
\multirow{4}{*}{PGD}     &  \textbf{Proposed Method}        &   \textbf{98.4}       &   \textbf{97.0}     &      \textbf{96.1}               &    \textbf{89.0}  & \textbf{98.0}  &  \textbf{97.5}    &    \textbf{99.3}                    &   \textbf{88.5}        &    \textbf{99.6}    \\\cline{2-11} 
&       Shaker et al. \cite{shaker2020generalization}    &   96.5      &    93.4     &   92.8          & 81.3   &   97.9               &  94.5              &    98.2             &  82.8 &   97.4     \\ \cline{2-11}
&         Kachuee et al. \cite{kachuee2018ecg}    &    87.2      &   74.0     &     86.9            & 66.3 &  87.6    &         84.7    &         91.8                 &    65.6         &    95.2  \\ \cline{2-11}
&        Acharya et al. \cite{acharya2017deep}     &  77.2     &   54.0    &    88.8         & 54.1 & 91.3     &  57.8            &      82.0                   &     66.2       &   80.6     \\ \hline\hline
\multirow{4}{*}{CW}      &   \textbf{Proposed Method}        &     \textbf{98.8}     &     \textbf{97.8}   &    \textbf{97.0}             &  \textbf{91.1}  & \textbf{98.8}    &    \textbf{98.3}        &   \textbf{99.5}                     &      \textbf{91.0}     &  99.5            \\\cline{2-11} 
&        Shaker et al. \cite{shaker2020generalization}    &   95.4       &   90.8      &    96.0          &  84.7 & 97.0      &         97.8     &           94.1                 &   72.6             &       \textbf{99.7}         \\ \cline{2-11} 
&         Kachuee et al. \cite{kachuee2018ecg}    &   91.9      &    84.5     &      83.3      &  74.4  & 89.3   &   79.7 & 99.3               &         61.1     &     96.3                 \\\cline{2-11}
&       Acharya et al. \cite{acharya2017deep}     &   81.2      &     61.8   &  89.1              & 64.6  & 81.5  &         67.7    &    94.1          &    83.4      &  86.8     \\ \hline\hline
\multirow{4}{*}{DBB}      &   \textbf{Proposed Method}        &     \textbf{93.0}     &     \textbf{85.8}   &    \textbf{94.9}            &  \textbf{84.4}  & \textbf{96.1} &    \textbf{91.3}           &   \textbf{96.2}             &    \textbf{84.1}            &  \textbf{93.9}            \\\cline{2-11} 
&        Shaker et al. \cite{shaker2020generalization}    &  90.1      &   80.6      &   86.7          &  65.1 & 94.2     &        79.4     &      94.9               &  83.4             &       91.7       \\ \cline{2-11} 
&         Kachuee et al. \cite{kachuee2018ecg}    &   79.7      &   57.9     &     86.4      &  78.0  & 69.3   &  70.9 & 96.6               &      82.1     &   93.6                \\\cline{2-11}
&       Acharya et al. \cite{acharya2017deep}     &   81.9      &  62.6   &   85.8           & 68.1  & 75.6   &       77.0    &    95.8           &   82.8      &  92.7      \\ \hline\hline
\multirow{4}{*}{HSJ}      &   \textbf{Proposed Method}        &     \textbf{71.9}     &     \textbf{44.8}   &    \textbf{68.2}            &  \textbf{34.6}  & \textbf{78.6}      &   35.7        &   \textbf{82.6}                 &      32.4     & 83.8          \\\cline{2-11} 
&        Shaker et al. \cite{shaker2020generalization}    &   70.5       &   41.2      &  67.4          &  33.0 & 77.1     &        \textbf{40.4}     &           81.1               & 37.5            &       83.4        \\ \cline{2-11} 
&         Kachuee et al. \cite{kachuee2018ecg}    &   68.6      &   39.3    &   65.8   &  21.3 & 82.5  &  10.3 & 94.5              &    \textbf{43.9}    &     62.2              \\\cline{2-11}
&       Acharya et al. \cite{acharya2017deep}     &  68.4     &   37.6   &  70.0             & 32.3  & 57.4 &  23.6     &    90.4           &    20.3     &  \textbf{90.0}     \\ \hline
\end{tabular}
\end{adjustbox}
\label{table1}
\end{table}

\subsection{Hyper-parameters}
We chose  $ \lambda_{atk} =10$ (Eq.~\ref{eq1}),  $ \lambda_{ary} =10$ (Eq.~\ref{eq2}), and $ \lambda_{mse} =1$ (Eq.~\ref{eq3}), to give more weight to classification losses than to adversarial loss. We give more weight to attacked signals than non-attacked ones by using $ \kappa=[1,1.05]$ (Eq.~\ref{eq1}). We used Adam optimizer \cite{kingma2014adam} with a learning rate of $\alpha=0.0001$, $\beta_1=0.5$ and $\beta_2=0.999$. We used Tensorflow 2.0 to train the model with batch size, $b=128$ for 100 epochs taking 4 hours to train on NVIDIA P100 GPU. We initialized the noise vector, $z$ with float values between $[0,1]$. Code repository is provided in this \href{https://github.com/FarihaHossain/ECG-ATK-GAN}{link}.

\subsection{Quantitative Evaluation}
We perform the quantitative evaluation by comparing our model with other state-of-the-art architectures \cite{shaker2020generalization,acharya2017deep,kachuee2018ecg} on both attacked and non-attacked data from MITBH and PTB datasets. In the first experiment, we use either only normal or adversarial attacked test data (19,327 and 4,000 for MITBH and PTB) for benchmarking the models on normal/abnormal beat classification, which is illustrated in Table.~\ref{table1} and Table.~\ref{table2}. We train all the models on their respective attacked and non-attacked training samples for a fair comparison. For metrics, we use Accuracy, Sensitivity, and Specificity. We can see that for \textbf{`No Attack'}, all models achieve comparatively good results. However, for each distinct attack, the results worsen for other models compared to ours. The architecture in \cite{shaker2020generalization,acharya2017deep,kachuee2018ecg} utilizes 1D Convolution based architecture. Out of these models, Shaker et al. \cite{shaker2020generalization} adopt DC-GAN, a generative network for adversarial signal generation. However, their classification architecture is trained separately, and they provide results only on real ECG signals. One reason for their model's good performance for the no-attack scenario is training with GAN-generated adversarial samples, which helps to learn out-of-distribution signals. Moreover, the two 1D CNN architectures achieve better sensitivity for minority category \textbf{F} for FGSM, BIM, and HSJ attacks. Similarly, our model's performance on the minority category \textbf{F} is best for PGD, CW, and DBB attacks and second best for FGSM and BIM. Our model performs poorly against HSJ attacks because the signals have too much high noise and no clear pattern, as illustrated in \ref{fig3}. Besides that, our architecture's overall performance is more robust against adversarial attacks for classifying arrhythmia and myocardial infractions, as shown in Table.~\ref{table2}.

\begin{table}[!tp]
\caption{\textbf{PTB Dataset} : Comparison of architectures trained and evaluated on \textbf{non-attacked/attacked} ECGs for normal and myocardial infarction beat classification.}
\centering
\begin{adjustbox}{width=0.6\textwidth}
\begin{tabular}{c||c||c|c|c|c|c|c|c}
\hline
                                          & \multicolumn{1}{c||}{\textbf{Methods}}                                          & \multicolumn{1}{c|}{\textbf{No Attack}} & \multicolumn{1}{c|}{\textbf{FGSM}} & \multicolumn{1}{c|}{\textbf{BIM}} & \multicolumn{1}{c|}{\textbf{PGD}}  & \multicolumn{1}{c|}{\textbf{CW}} & \multicolumn{1}{c|}{\textbf{DBB}} & \multicolumn{1}{c}{\textbf{HSJ}} \\ \hline
\multicolumn{1}{r||}{\multirow{4}{*}{Accuracy}} & \textbf{Proposed Method}                                       & \multicolumn{1}{c|}{\textbf{99.5}}                   &      \textbf{99.4}                              &            \textbf{99.6}                                      &          \textbf{99.6}                         &            \textbf{99.5}                      &           \textbf{93.1}                        &           \textbf{71.8}                       \\ \cline{2-9} 
\multicolumn{1}{r||}{}                     & Shaker et al. \cite{shaker2020generalization} & \multicolumn{1}{c|}{98.0}                   &                            98.6        &        95.8                           &          96.4                           &          98.3                       &                 91.4                 &      70.2                                                            \\ \cline{2-9} 
\multicolumn{1}{r||}{}                     & Kachuee et al. \cite{kachuee2018ecg}          & \multicolumn{1}{c|}{95.2}                   &              97.1                      &             94.4                     &              92.2                      &                       91.3            &               88.6                &       56.5                                                          \\ \cline{2-9} 
\multicolumn{1}{r||}{}                     & Acharya et al. \cite{acharya2017deep}         &      79.8                                 &           84.1                           &           83.2                        &         84.1                            &         80.9                          &                77.4                  &          54.7                                                    \\ \hline\hline
\multirow{4}{*}{Sensitivity}                      & \textbf{Proposed Method}                                       &         
\textbf{99.3}                              &           \textbf{99.2}                           &                 \textbf{99.6}                  &     \textbf{99.7}                                                           &                  \textbf{99.2}                &              \textbf{92.6}                     &             79.8                  \\ \cline{2-9}
                                          & Shaker et al. \cite{shaker2020generalization} &          96.7                             &          98.3                          &            92.1                       &                    94.8                &         98.0                         &                      91.5            &       83.7                                                         \\ \cline{2-9} 
                                          & Kachuee et al. \cite{kachuee2018ecg}          &     98.3                                    &           96.0                         &     95.9                              &                     92.1              &      95.4                             &                    86.7            &          85.5                                                       \\ \cline{2-9} 
                                          & Acharya et al. \cite{acharya2017deep}         &                82.1                       &            93.1                          &      93.4                             &               90.6                     &                       90.3            &                 88.5                 &         \textbf{90.7}                                                         \\ \hline\hline
\multirow{4}{*}{Specificity}                      & \textbf{Proposed Method}                                       &          \textbf{99.7}                             &         \textbf{99.5}                             &                 \textbf{99.7}                  &        \textbf{99.5}                           &               \textbf{99.7}                    &               \textbf{93.7}                      &                    \textbf{64.0}                                              \\ \cline{2-9} 
                                          & Shaker et al. \cite{shaker2020generalization} &         99.3                                &     98.9                               &               99.4                    &                 98.0                   &        98.7                          &                   91.2               &      56.8                                                        \\ \cline{2-9} 
                                          & Kachuee et al. \cite{kachuee2018ecg}          &     92.1                                    &          98.2                          &      93.0                             &                  92.2                 &         87.3                          &                     90.3             &             28.1                                                   \\ \cline{2-9} 
                                          & Acharya et al. \cite{acharya2017deep}         &        77.7                                 &        75.3                            &      73.2                             &             77.8                   &                        71.6           &                           66.5       &                                                       19.4         \\ \hline
\end{tabular}
\end{adjustbox}
\label{table2}
\end{table}

\begin{table}[!tp]
\caption{\textbf{Generator's Performance}: Similarity of adversarial and attacked / non-attacked signals.}
\centering
\begin{adjustbox}{width=1\textwidth}
\begin{tabular}{c|cccc||cccc}
\hline
\multicolumn{1}{c|}{} & \multicolumn{4}{c||}{\textbf{MITBIH}}                                                                                                                                                                                                                                                          & \multicolumn{4}{c}{\textbf{PTB}}                                                                                                                                                                                                                                                              \\ \hline
\multicolumn{1}{l|}{} & \multicolumn{1}{c|}{Mean-Squared-Error} & \multicolumn{1}{c|}{\begin{tabular}[c]{@{}c@{}}Structural\\ Similarity\end{tabular}} & \multicolumn{1}{c|}{\begin{tabular}[c]{@{}c@{}}Cross-corelation\\ Coefficiet\end{tabular}} & \multicolumn{1}{c||}{\begin{tabular}[c]{@{}c@{}}Normalized\\ RMSE\end{tabular}} & \multicolumn{1}{c|}{Mean-Squared-Error} & \multicolumn{1}{c|}{\begin{tabular}[c]{@{}c@{}}Structural\\ Similarity\end{tabular}} & \multicolumn{1}{c|}{\begin{tabular}[c]{@{}c@{}}Cross-corelation\\ Coefficient\end{tabular}} & \multicolumn{1}{c}{\begin{tabular}[c]{@{}c@{}}Normalized\\ RMSE\end{tabular}} \\ \hline
No Attack             & \multicolumn{1}{c|}{0.0129}    & \multicolumn{1}{c|}{99.90}                                                                & \multicolumn{1}{c|}{99.86}                                                                      &            3.487e-5                                                                    & \multicolumn{1}{c|}{0.0184}    & \multicolumn{1}{c|}{99.87}                                                                & \multicolumn{1}{c|}{99.93}                                                                       &               8.152e-5                                                                \\ \hline
FGSM                  & \multicolumn{1}{c|}{0.0117}    & \multicolumn{1}{c|}{99.81}                                                                & \multicolumn{1}{c|}{99.89}                                                                      &            2.890e-5                                                                    & \multicolumn{1}{c|}{0.0001}    & \multicolumn{1}{c|}{99.84}                                                                & \multicolumn{1}{c|}{99.89}                                                                       &            0.02391                                                                   \\ \hline
BIM                   & \multicolumn{1}{c|}{0.0134}    & \multicolumn{1}{c|}{99.80}                                                                & \multicolumn{1}{c|}{99.86}                                                                      &           3.737e-5                                                                     & \multicolumn{1}{c|}{0.0155}    & \multicolumn{1}{c|}{99.91}                                                                & \multicolumn{1}{c|}{99.95}                                                                       &             5.769e-5                                                                  \\ \hline
PGD                   & \multicolumn{1}{c|}{0.0122}    & \multicolumn{1}{c|}{99.84}                                                                & \multicolumn{1}{c|}{99.89}                                                                      &               3.115e-5                                                                 & \multicolumn{1}{c|}{0.0179}    & \multicolumn{1}{c|}{99.88}                                                                & \multicolumn{1}{c|}{99.92}                                                                       &                                   7.722e-5                                            \\ \hline
CW                    & \multicolumn{1}{c|}{0.0065}    & \multicolumn{1}{c|}{99.95}        & \multicolumn{1}{c|}{99.97}                                                                      &                  9.038e-6                                                              & \multicolumn{1}{c|}{0.0188}    & \multicolumn{1}{c|}{99.87}                                                                & \multicolumn{1}{c|}{99.92}                                                                       &          8.498e-5                                                                     \\ \hline
DBB                   & \multicolumn{1}{c|}{0.0002}    & \multicolumn{1}{c|}{99.00}                                                                & \multicolumn{1}{c|}{99.39}                                                                      &                              0.03159                                                  & \multicolumn{1}{c|}{0.0007}    & \multicolumn{1}{c|}{99.19}                                                                & \multicolumn{1}{c|}{99.29}                                                                       &                                   0.05532                                            \\ \hline
HSJ                   & \multicolumn{1}{c|}{0.0003}    & \multicolumn{1}{c|}{99.42}                                                                & \multicolumn{1}{c|}{99.45}                                                                      &                              0.0393                                                  & \multicolumn{1}{c|}{0.0003}    & \multicolumn{1}{c|}{99.51}                                                                & \multicolumn{1}{c|}{99.60}                                                                       &                                                 0.03872                              \\ \hline
\end{tabular}
\end{adjustbox}
\label{table3}
\end{table}

\subsection{Qualitative Evaluation}
% ***********Intra Patient************
For finding the similarity between real and synthesized attacked/non-attacked ECG signals, we benchmarked generated adversarial signals using four different metrics, i) Mean Squared Error (MSE), ii) Structural Similarity (SSIM), iii) Cross-correlation coefficient, and iv) Normalized Mean Squared Error (NRMSE). In Table.~\ref{table3}, We use both attacked and non-attacked signals from the test set.  We score SSIM of $99.90\%$, $99.81\%$ (FGSM), $99.80\%$ (BIM), $99.84\%$ (PGD), $99.95\%$ (CW), $99.00\%$ (DBB) and $99.43\%$ (DBB) for MITBH Dataset. On the other hand we achieve SSIM of $99.87\%$ (No Attack), $99.84\%$ (FGSM), $99.91\%$ (BIM), $99.84\%$ (PGD), $99.87\%$ (CW), $99.19\%$ (DBB) and $99.51\%$ (DBB) for PTB Dataset. As for cross-correlation, MSE, and NRMSE, our model generates quite realistic signals with minimal error.

% ***********Benchmarking************

\section{Conclusions and Future Work}
\label{sec:conclution}
This paper presents ECG-ATK-GAN, a novel conditional Generative Adversarial Network for accurately predicting different types of arrhythmia from both regular and adversarially attacked ECGs. In addition, our architecture incorporates a new class-weighted categorical objective function for capturing out-of-distribution signals and robustly discerning class-specific features corrupted by adversarial perturbations. We provided an extensive benchmark on two publicly available datasets to prove the robustness of our proposed architecture. One future direction is to improve our architecture by defending against other types of adversarial attacks.

\textbf{Prospect of application:} Detecting arrhythmia accurately and robustly in real-time will pave the way for better patient care and disease monitoring. In addition, insurance companies, contractors, partners, and many stakeholders will financially benefit from a trusted cardiac arrhythmia diagnostic system that is robust against adversarial attacks. This system can also help identify new attack types by distinguishing signal anomalies.

\bibliographystyle{splncs04}
% \bibliography{mybibliography}
%
\bibliography{reference}

\iffalse

\fi

\end{document}